# Turbulence in the Ocean, Atmosphere, Galaxy, and Universe

Carl H. Gibson

*Departments of Applied Mechanics and Engineering Sciences*
*and Scripps Institution of Oceanography*
*University of California at San Diego*
*La Jolla, CA 92093-0411*

## Abstract

Flows in natural bodies of fluid often become turbulent, with eddy-like motions dominated by inertial-vortex forces. Buoyancy, Coriolis, viscous, self-gravitational, electromagnetic, and other force constraints produce a complex phase space of wave-like hydrodynamic states that interact with turbulence eddies, masquerade as turbulence, and preserve information about previous hydrodynamic states as fossil turbulence. Evidence from the ocean, atmosphere, galaxy and universe are compared with universal similarity hypotheses of Kolmogorov (1941, 1962) for turbulence velocity u, and extensions to scalar fields like temperature mixed by turbulence. Universal u and spectra of natural flows can be inferred from laboratory and computer simulations with satisfactory accuracy, but higher order spectra and the intermittency constant $\mu$ of the third Kolmogorov hypothesis (1962) require measurements at the much larger Reynolds numbers found only in nature. Information about previous hydrodynamic states is preserved by Schwarz viscous and turbulence lengths and masses of self-gravitating condensates (rarely by the classical Jeans length and mass), as it is by Ozmidov, Hopfinger and Fernando scales in hydrophysical fields of the ocean and atmosphere. Viscous-gravitational formation occurred $10^4$-$10^5$ y after the Big Bang for supercluster, cluster, and then galaxy masses of the plasma, producing the first turbulence. Condensation after plasma neutralization of the H-$^4$He gas was to a primordial fog of sub-solar particles that persists today in galactic halos as "dark matter". These gradually formed all stars, star clusters, etc. (humans!) within.

## 1. Introduction

This paper provides a brief review of turbulence in large bodies of natural fluids such as the ocean, atmosphere, galaxy and universe. Ratios of the space and time scales of the largest and smallest eddy motions in these fluids may be much larger than those observed in laboratory flows or simulated in a computer. Ratio sizes are determined by the Reynolds, Froude, Rossby, and other dimensionless groups of the flows that compare the inertial vortex forces of eddy motions to viscous, buoyancy, Coriolis, gravitational, electro-magnetic and other forces that arise to damp out the turbulent eddies. Experimental evidence about turbulence in natural flows is difficult to acquire since space and time scales of the eddies may be inconveniently large in the ocean and atmosphere and impossibly large in galaxies and the extra-galactic universe, often exceeding times available for individual investigations (and human lifetimes) by large factors. A survey of the literature shows that the rate of papers written about turbulence decreases as the eddy time and length scales increase. Thousands of papers per year are published about laboratory and atmospheric turbulence, hundreds about oceanic turbulence, dozens about galactic turbulence, but only a handful about turbulence in the intergalactic medium. None were found about the primordial transition to turbulence. Much remains to be learned about turbulence with the very many degrees of freedom in both damping and generation modes and huge ranges of space and time scales found in natural flows. It is particularly important, necessary, and difficult, to define which flows should be considered turbulent and which should not.

Unfortunately, fine distinctions between flow classes are often not drawn in the various literatures of natural flows. Any random flow or scalar fluctuation may be referred to as "turbulence". Stars are said to twinkle because the atmosphere is "turbulent" although stars actually twinkle because random refractive index fluctuations exist in the line of sight, and these may or may not have been produced by turbulence, either past or present. In the ocean, turbulence is extremely rare, difficult to sample, and rapidly fossilized to remnant wave motions or microstructure. Correct interpretation of the measurements is complicated by buoyancy and Coriolis effects. A vigorous debate has gone on in oceanography for over two decades about whether or not turbulence in the ocean interior dominates the vertical diffusion of heat and dissolved chemical species. When the sampled ocean microstructure is treated as turbulence rather than fossils of turbulence, one reaches the latter conclusion: that the "turbulence" patches detected are too rare and the dissipation rates of kinetic energy and species variance in these patches are too low to account for the necessary fluxes. How the ocean mixes is left as a mystery.

An alternative interpretation, Gibson (1982), is that the turbulence diffusion process has been undersampled and underestimated. The largest and most powerful patches sampled are invariably fossil turbulence at the largest scales; that is, remnant fluctuations of temperature, density and salinity formed by turbulence in fluid that is no longer actively turbulent at the scale of the fluctuations, Gibson (1980, 1996cd). Think of jet contrails. Previously overturning turbulent eddies are converted by buoyancy forces to bobbing motions at the ambient stratification frequency N or less. These motions constitute a unique class of initially saturated internal waves termed fossil vorticity turbulence, Gibson (1986). Laboratory measurements in stratified wakes demonstrate the phenomenon, Stillinger et al. (1983), Xu et al. (1995). The uniqueness and persistence of the mixing patterns of turbulence permit construction of "hydrodynamic phase diagrams", or "turbulence activity diagrams", to classify oceanic and lake microstructure according to hydrodynamic state, Gibson (1980, 1986, 1990, 1991abcd), Imberger and Ivey (1991). Phase diagrams also apply to anisotropic turbulence constrained in the vertical by buoyancy forces and in the horizontal by Coriolis forces, producing "2D fossil turbulence", Gibson (1990, 1991cd). The undersampling interpretation indicated by the fossils is verified statistically by taking the extreme intermittency of the viscous and temperature dissipation rates into account when estimating averages and vertical heat fluxes, Gibson (1991b).

Similar processes occur for atmospheric turbulence (clouds, jet contrails). However, atmospheric density and temperature microstructure patches in stratified layers are rarely sampled to the molecular scales required for classification so their hydrodynamic states are usually unknown. Information about stellar, interstellar, galactic, and extragalactic scale turbulence and fossil turbulence is even more fragmentary. The fossil turbulence concept was independently (and perhaps first) introduced by Gamov (1954), who suggested [p. 483] "the present distribution of galaxies in space represents a 'fossilized turbulence' in the [primordial] gas, immediatelypreceding the formation of galaxies". Even though Gamov's assumptions of strong primordial turbulence triggering Jeans instability were erroneous, he was correct in assuming that many self-gravitational structures of cosmology and astrophysics, including stars, globular star clusters, galaxies, and clusters and superclusters of galaxies, reflect the hydrodynamic state of the fluid at their times of formation and are indeed fossils of primordial turbulence and non-turbulence.

In the present review, turbulence is defined narrowly as an eddy-like state of fluid motion where the eddies are dominated by the inertial-vortex force per unit mass $\vec{v} \times \vec{\omega}$, where $\vec{v}$ is the fluid velocity and $\vec{\omega}$ = curl $\vec{v}$ is the vorticity. This non-linear term in the conservation of momentum equations causes vortex sheets (shear layers) to be unstable and break up into eddies, greatly enhancing the transport of any hydrophysical properties of the fluid such as chemical species concentration, sensible heat, vorticity, or momentum itself compared to molecular diffusion, and scrambling any large scale organized structures. Such flows are sometimes termed "true turbulence", "strong turbulence", "Kolmogorovian turbulence", or simply "fluid mechanical turbulence", versus "plasma turbulence", "geophysical turbulence", "quantum turbulence" and other flows that may have properties similar to those of turbulence such as random or nonlinear behavior, but are dominated by other forces. Note that irrotational flows cannot be turbulent by this definition.

When they occur, the eddy motions of turbulence grow in natural bodies of fluid until they are constrained by one of the other possible forces. Turbulence is then damped out or converted to another class of fluid motion that may preserve information about the previous turbulence as fossil turbulence. In the ocean and atmosphere the flow is usually not turbulent at any particular place and time at any scale, by our definition, because the viscous, surface tension, buoyancy and other forces generally dominate the $\vec{v} \times \vec{}$ forces. Nevertheless, these rare turbulence events cannot be ignored in vertical transport processes because molecular diffusion is so enormously slow. The thermal time constant $h^2/$ of the ocean is about a million years and the salinity time constant $h^2/D$ is about a billion years, where h is the depth and    and D are the molecular diffusivities, yet a hundred years of hydrography show that such passive ocean properties are somehow, somewhere, mixed much more rapidly, with estimated "residence times"    $M/\dot{M}$ of order a thousand years, where M is the ocean mass and $\dot{M}$ is the rate of introduction of water masses with different temperatures and salinities. Diffusion of man-made destructive catalytic gases up into the ozone layer would take hundreds of millions of years without the assistance of rare bursts of turbulence in the stratified atmosphere below.

Turbulence in the sun is strongly inhibited by magnetic, Coriolis, and gravitational forces, but is not prevented. The time constant for changes in the global magnetic field of the sun $L^2/$ would be of order a million years based on the size of the sun L    $10^9$ m and the magnetic diffusivity    $10^5$ m$^2$ s$^{-1}$. The actual magnetic cycle time is only 22 years because the magnetic diffusivity is enhanced several orders of magnitude by turbulence. Thermonuclear pressure cookers of heavy atoms in the cores of stars are controlled by the turbulence diffusion of heat and mass through stratified layers. The turbulence, as usual, is subject to powerful gravitational, magnetic, and Coriolis constraints. Radial transport in accretion disks of stars and black holes are determined by such constrained turbulence diffusion. Differential velocities measured in galactic disks, Rubin (1984), are generally practically nil outside the inner $10^{20}$ m core, contrary to Kepler's laws of planetary orbital velocities that require a decrease as the radius $r^{-1/2}$. Possible mechanisms to smooth out the velocity gradients include turbulence diffusion and a complex interaction of magnetic, gravitational, and Coriolis waves. Constant velocity requires an increase in centrifugal forces with radius, so unobserved dark matter in the galaxy halo is required to bind the galactic mass together by gravity.

Even though the evidence in natural flows is sparse, it consistently supports the same conclusion: turbulence processes are crucial to the evolution of galaxies, stars and planets (and planetary oceans and atmospheres). Turbulence has not always existed. According to a new model of self-gravitational structure formation based on viscous and turbulence forces of real fluids, Gibson (1996a), turbulence first began gently, several thousand years after the Big Bang, driven by the first self-gravitational, viscosity-limited, condensations of hot viscous plasma at the scale of causal connection. The classical Jeans (1902, 1929) theory for the acoustic condensation of ideal fluids is questioned as an adequate model for structure formation in cosmology and astrophysics. The new model gives a very different prediction for structure formation in the universe than predictions from the Jeans model alone.  In the following, we review turbulence fundamentals, apply the results to the ocean, atmosphere, galaxy, and universe, and summarize.

## 2.  Turbulence fundamentals

### 2.1.  The turbulence cascade and universal similarity

Turbulent flows can be described by conservation of momentum equations for fluids in the form

$$\frac{\partial \vec{v}}{\partial t} = -\nabla B + \vec{v} \times \vec{\omega} + \vec{F}_b + \vec{F}_c + \vec{F}_v + \vec{F}_m + \vec{F}_{etc.} \tag{1}$$

where $\vec{v}$ is the velocity, $\vec{\omega}$ is the vorticity or curl $\vec{v}$, B is the Bernoulli group $p/\rho + v^2/2 + gz$, g is gravity, z is up, and $\vec{F}_b + \vec{F}_c + \vec{F}_v + \vec{F}_m + \vec{F}_{etc.}$ are the buoyancy, Coriolis, viscous, magnetic, etc. body forces that arise to damp out turbulent eddies resulting from inertial vortex forces $\vec{v} \times \vec{\omega}$. Flow regimes are classified according to dimensionless groups such as the Reynolds number Re $(\vec{v} \times \vec{\omega})/\vec{F}_v$, the Froude number Fr $(\vec{v} \times \vec{\omega})/\vec{F}_b$ , and the Rossby number Ro $(\vec{v} \times \vec{\omega})/\vec{F}_c$.  Taking the curl of (1) gives the vorticity equation

$$\frac{D\vec{\omega}}{Dt} = \vec{\omega} \cdot \overleftrightarrow{e} + \nabla \times (\vec{F}_b + \vec{F}_c + \vec{F}_v + \vec{F}_m + \vec{F}_{etc.}) \tag{2}$$

where $\overleftrightarrow{e}$ is the rate of strain tensor with components $e_{ij} = (u_{i,j} + u_{j,i})/2$.  We see from (2) that the vorticity of a fluid particle on the left hand side can change only due to local vortex stretching in the first term on the right, or curls of the various forces in the second term on the right which are also local.

Turbulence eddies usually develop first at small (viscous) scales on surfaces in the fluid where $\vec{\omega}$ is large (shear layers).  Such vortex sheets are unstable at all scales L, but the eddy turnover time $T(L) = L/V(L)$ is smallest for the smaller scales unless the perturbation velocity $V(L)$ is forced to be very large at some particular large scale; for example, wing tip vortices forced at the scale of the wing W.  These have time scales W/U that can be quite small since U is the speed of the aircraft, although not smaller than the Kolmogorov scale of the wing boundary layers which are proportional to $U^{-3/2}$.  The wingtip vortex spectra will initially have a spectral gap to fill in between wave numbers k near 1/W and $1/5L_K$ as eddies cascade up from $5L_K$ to W scales.

Mixing layers of thickness h and velocity difference $U_o$ form eddies on scales proportional to h, even if h is larger than $5L_K$ (the scale of the universal critical Reynolds number).  This is because the overturn time $y/u = h/U_o$ is the same for all scales $y < h$, since $u(y) \approx yU_o/h$ in the mixing layer.  However, in order to form a mixing layer in the first place requires a sudden action such as a wind tunnel contraction, the high speed passage of a wing, or the rapid expulsion of fluid past a separation point to form a starting vortex or smoke ring.  These flows form vortex structures on scales larger than $5L_K$ without forming eddies at all intermediate scales in the process. For a mixed layer, $h = L_K Re_h^{1/2}$, where $Re_h = U_o h/\nu$ and $L_K \approx (\nu h/U_o)^{1/2}$. If the mixed layer Reynolds number $Re_h$ is larger than the universal critical value of about $Re_{crit} \approx$ 25-100, the mixing layer will become turbulent during any gradual formation process with a universal spectrum in quasi-equilibrium typical of most turbulent flows.

We can estimate the probable dependence of $V(L)$ from the Kolmogorov (1941) second universal similarity hypothesis for the probability law $F_n$ for velocity differences between n points in space separated by n vectors $\vec{y_k}$, $1 \leq k \leq n$. According to the hypothesis, turbulent velocity fields for very large Re values should be homogeneous, isotropic, and depend only on the viscous dissipation rate $\varepsilon$ and the magnitude of the separation vectors $y_k$. The magnitudes of $y_k$ must all be larger than the viscous (Kolmogorov) length scale $L_K = (\nu^3/\varepsilon)^{1/4}$ and smaller than the energy (Obukhov) length scale $L_O$, where $\nu$ is the kinematic viscosity. If the flow surrounding the vortex sheet is forced by velocity with a spectrum like that of turbulence, the perturbation velocity $V(L)$ should depend only on $\varepsilon$ and L according to the Kolmogorov second hypothesis, so

$$V(L) \approx \varepsilon^{1/3} L^{1/3} \tag{3}$$

and

$$T(L) = L/V(L) \approx L^{2/3} \varepsilon^{-1/3}, \quad 5L_K < L < L_O \tag{4}$$

which shows the smaller eddies form first since their overturn times decrease as $L^{2/3}$. The spectrum of the perturbation velocity must be steeper than $k^{-3}$ between wavenumber $k = L^{-1}$ and $k = L_K^{-1}$ for any eddy larger than $5L_K$ to form in smaller times than those at the Kolmogorov scale, where k is the wavenumber. Otherwise eddies form first at the Kolmogorov scale. These pair with other eddies, the pairs of eddies pair with other pairs, building up larger and larger values of the Obukhov scale $L_O$ as the size of the turbulence region grows.

The process is illustrated by jet, boundary layer, and wake turbulent flows in Figure 1. The turbulent boundary layer is described by the "law of the wall", where $u^+ = u/u^*$ and $y^+ = y/y^*$ are equal in the viscous sublayer for $u^+ = y^+ \leq 5$, and $u^+ = (1/\kappa) \ln (y/y^*) + B$ for $y^+ \geq 5$, where $\kappa$ and B are universal constants. The friction length $y^* = \nu/u^*$ equals the Kolmogorov length scale $L_K = (\nu^3/\varepsilon)^{1/4}$ and the friction velocity $u^* = (\tau/\rho)^{1/2}$ equals the Kolmogorov velocity scale $(\nu\varepsilon)^{1/4}$, $\nu$ is the kinematic viscosity and $\varepsilon$ is the viscous dissipation rate near the wall. The transition to turbulence at $u^+ = y^+ = 5$ provides a measure of the universal critical Reynolds number $Re_{crit} \approx$ 25, since $Re_{trans.} = (uy/\nu)_{crit.} = 25 (u^*y^*/\nu)_{crit.} = 25 (u_K y_K/\nu)_{crit.} = 25$, and $Re_{trans.}$ should equal $Re_{crit}$ according to the first universal similarity hypothesis of Kolmogorov (1941). By hypothesis, transition to turbulence should occur at approximately $5L_K$ in all turbulence, as shown in the examples of turbulence in Fig. 1. By Kolmogorov's second hypothesis, turbulence is turbulence, meaning that all the flows of Fig. 1 should develop with complete geometrical, kinetic, and dynamic similarity whether the jets, wakes and boundary layers are on cosmic or laboratory scales. For second order statistical parameters like power spectra of turbulent velocity fluctuations and passive scalar fields like temperature, the Kolmogorov (1941) hypotheses and their scalar field counterparts have withstood exhaustive experimental scrutiny and proved valid, Gibson (1991a).

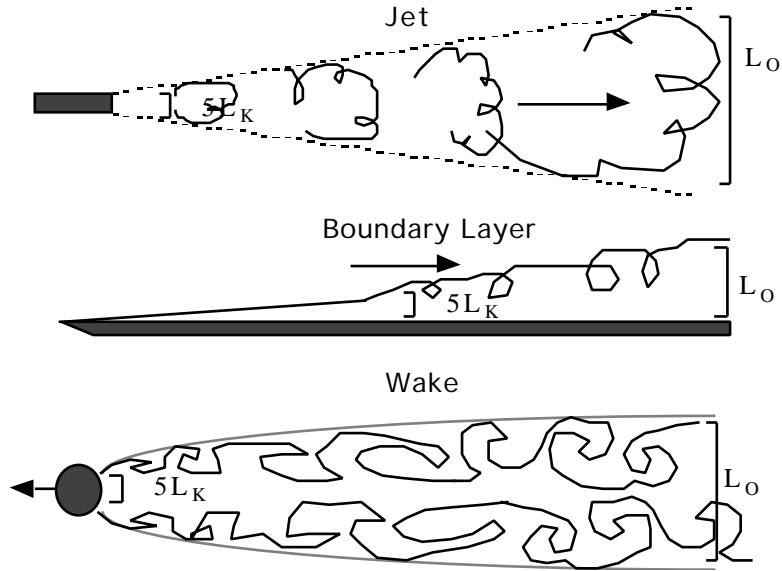

**Figure 1.** Examples of turbulence kinetic energy formation and evolution. Initially $5L_K$ $L_O$, where $L_K$ is the (viscous) Kolmogorov scale and $L_O$ is the (energy) Obukhov scale. Finally, $L_O \gg 5L_K$, after a cascade from small scales to large by processes of eddy pairing and entrainment of (nonturbulent) irrotational fluid.

The third hypothesis of Kolmogorov (1962) proposes a refinement of the first two hypotheses to take into account the fact that    is not a constant, but an intermittent lognormal random variable with intermittency factor, or variance of the natural logarithm

$$^2_{\ln\ r} = \mu \ln (L_O/r) \qquad (5)$$

that increases logarithmically with decreasing averaging scale r for a given energy scale $L_O$, where $\mu$ is a universal constant for very large Reynolds number turbulent flows. This refinement has very slight effects on statistics like power spectra of u and scalar fields  , but major effects on spectra of their dissipation rates and other higher order statistics of turbulence and turbulent mixing. The appropriate value of $\mu$ for natural flows is important, since (5) provides a means of inferring macroscales $L_O$ of a flow from measurements of the intermittency factor $^2_{\ln\ r}$, but is currently a subject of vigorous debate. The difference between values of $\mu$    0.16-0.3 characteristic of estimates from laboratory flows and numerical simulations and the values of $\mu$ 0.4-0.5 characteristic of natural flows is related to the evolution of different classes of geometrical singularities of the multifractal dissipation field networks and their degeneration as the Reynolds number of the flows becomes asymptotically large, where $\mu$     1/2 as shown by Bershadskii and Gibson (1994). Estimates of $L_O = r \exp(\sigma^2_{\ln r}/\mu)$ are extremely sensitive to the value of $\mu$; for example, if $\sigma^2_{\ln r}$ is 5 for r = 1 meter typical of the ocean seasonal thermocline scrambled horizontally by a storm, then the macroscale of turbulence is 100 km for $\mu$ = 0.43, which is realistic. However, $L_O$ is more than $10^{10}$ km (!) for $\mu$ = 0.16, which is not realistic for a cascade that should at least fit on the earth's surface with circumference $4 \times 10^7$ meters.

A common misconception is that turbulence generally forms by a cascade from large scales to small according to the Richardson poem, "Big whorls have smaller whorls, that feed on their velocity. And smaller whorls, have smaller whorls, and so on to viscosity (in the molecular sense)". The three examples of familiar turbulence flows in Figure 1 show that something is wrong with this idea. In each case turbulence begins at small scales and grows larger: not the other way around. The turbulence grows from small scales to large in what would commonly be termed an "inverse cascade". This "normal" turbulence cascade direction from large to small is true only if, by definition, the term "turbulence" includes the irrotational flow of external fluid into the interstices of the turbulent eddies as it is entrained. However, such inviscid, irrotational flows (ideal flows) are very different from flows that are dominated by inertial vortex forces (turbulence). Kinetic energy length scales are reduced from large to small as this irrotational external fluid is entrained at sizes larger than $L_O$, as shown schematically in Figure 2.

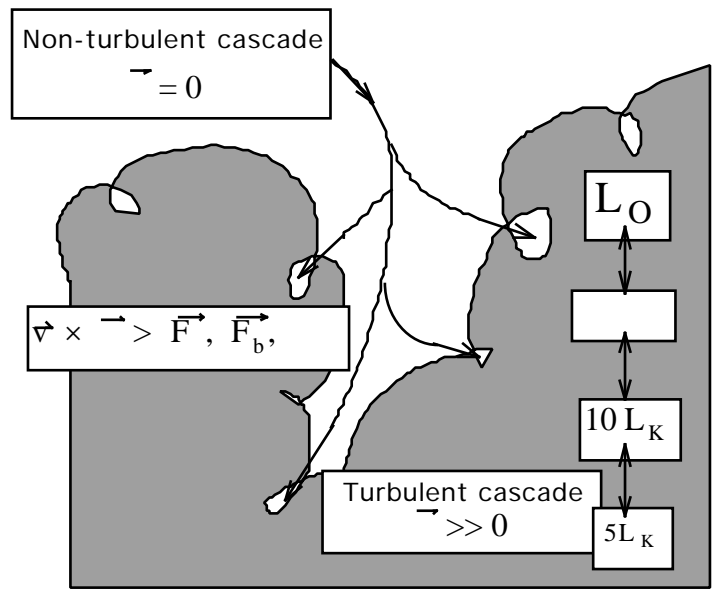

**Figure 2.** Energy cascades. Non-turbulent kinetic energy cascade in irrotational fluid from large scales to small, entrainment at viscous scales $5L_K$, with turbulence energy cascade from small scales to large (with feedback).

Misconceptions have negative consequences. Cosmology models have sometimes suggested that a long period of time should be required before the primordial big eddies of the first turbulence could break down to small enough scales to be relevant to the formation of stars and galaxies. Lumley (1992) provides a mathematical model for estimating the value of such cascade times. The conclusion that turbulence was not important to early structure formation in the universe is correct, but for different reasons. For thousands of years after the Big Bang beginning there was no turbulence because the Reynolds number of the plasma was too small. Huge bulk viscosity values ( $10^{59}$ kg m$^{-1}$ s$^{-1}$) are estimated for the initial stages ($10^{-33}$ to 24 s) by Brevick and Heen (1994) to account for the huge entropy measured from the cosmic background radiation. Based on these values, Gibson (1996a) estimates that the initial Reynolds number of the universe was $10^{-30}$ (after inflation, at t = $10^{-33}$ s). Re increased to critical values only after a few thousand years of expansion and cooling, with $10^{28}$ m$^2$ s$^{-1}$ at t = $10^{11}$ to $10^{12}$ s inferred from the largest observed mass of superclusters; that is, $10^{47}$ kg according to Kolb and Turner (1994). Turbulence was initially not present at any scale, and when it first formed due to the viscous-gravitational collapse of plasma at the Hubble scale, it was very gentle and rapidly damped by photon viscosity. This is revealed by many fossils of non-turbulence in the oldest stars, globular clusters of stars, and elliptical galaxies, and by the extreme isotropy of the cosmic background radiation, with T/T values of only $10^{-5}$ to $10^{-6}$, where T is the uniform temperature of 2.726 K.

Turbulence began to dominate gravitational structure formation only in the thermonuclear stages millions and billions of years later. Formation of planets, stars and active galactic nuclei are crucially dependent on the complex interactions of turbulence, magnetic forces, and gravity. Turbulence inhibits star formation by mixing away density concentrations that would otherwise initiate gravitational collapse. Once collapse has begun, turbulence diffusion of mass may increase the growth rate of the collapsing body. Turbulence "viscosity" enhances the diffusion of angular momentum radially outward and mass radially inward in accretion disks to overcome centrifugal forces that might otherwise inhibit or prevent the gravitational collapse process just as turbulence can either prevent or enhance combustion processes.

Modern technological advances are vastly accelerating the availability of turbulence information about natural fluids. Satellites monitor motions of the oceans and atmosphere. The Hubble Space Telescope is expanding the visible universe at rates far beyond the speed of light in every direction. Personal computers of individual scientists can store and process information at rates available only to national research centers a few decades ago. Progress in understanding natural turbulence is becoming much less limited by the tools of information gathering and analysis, and depends more on the ability of investigators to synthesize the flood of results.

## 2.2. Approximation theories of turbulence

Turbulence remains an unsolved problem of classical physics. Numerous attempts to extrapolate analytical methods that have been quite successful in other branches of physics to solve the turbulence problem have failed. The many body gravitational interaction problem of Newtonian mechanics is also unsolved. Solutions are not closed for motions, past or future, of more than three bodies. For many interacting stars in a galaxy, or many galaxies in a cluster, the computations are so daunting that astrophysicists often prefer to represent such systems by the "hydrodynamics" of cosmic fluids, Binney and Tremaine (1987, p. 195).

For the analysis of interacting charged particles, excellent results have been achieved using perturbation methods and the so-called Dyson equations, Fynmann diagrams, and renormalization group techniques of quantum electrodynamics. Extrapolation of these methods to solve turbulence problems have had some success. For the "weak turbulence" waves of plasma physics the methods work well, Kadomsev (1965), but so far they have had mixed success for the "strong turbulence" of hydrodynamics. Typically the mathematical formalisms are extremely complex, rather non-physical, and without predictions that can be readily tested by experiment. The most determined effort over many years has been that of Robert Kraichnan and his collaborators. The initial steps are discussed by Monin and Yaglom (1975, p308). Kraichnan (1959) introduced the direct interaction approximation (DIA) among Fourier elements of a turbulent velocity field, emphasizing integrals over triads of "directly interacting" wavenumber vectors and neglecting higher order interactions. The theory predicts an inertial subrange of the turbulent kinetic energy spectrum $E(k) \sim (u_o \varepsilon)^{1/2} k^{-3/2}$, where $u_o$ is the integral velocity scale, $\varepsilon$ is the average viscous dissipation rate, and $k$ is the wavenumber magnitude, compared to the prediction based on Kolmogorov's second hypothesis that $E(k) \sim \varepsilon^{2/3} k^{-5/3}$, without any coupling at small scales with $u_o$, which is dominated by the larger scales. The Kolmogorov theory seems a better physical representation of the nearly decoupled cascade of self similar eddy motions described above, where the smallest scale eddies should become more and more independent of the largest ones as the Reynolds number of the flow increases. Moreover, the data for high Reynolds number spectra are in excellent agreement with a -5/3 power law, and inconsistent with -3/2.

However, for the spectrum of the dissipation $E_\varepsilon$ with integral $\varepsilon^2$ the Kraichnan approximation is more successful than that of Kolmogorov. Dimensional analysis gives $E_\varepsilon \sim (\varepsilon)^2 k^{-1}$ using Kolmogorov variables $\varepsilon$ and $k$, contrary to observations. However, $E_\varepsilon \sim (u_o \varepsilon)^{3/2} k^{1/2}$ using Kraichnan's variables $u_o$ and $k$, which is in excellent agreement with turbulent boundary layer measurements at very high Reynolds number of Gibson, Stegen and McConnell (1970) and Gibson and Masiello (1972) over the ocean, Wyngaard and Pao (1972) over land, and many others that $E \sim k^{-1/2}$.  From Kolmogorov's third hypothesis of (5) applied to the dissipation spectrum,

$$E \sim k^{-1+\mu} \qquad (6)$$

confirming indications of the highest available Reynolds number measurements, Gibson (1991a), that the asymptotic value of the Kolmogorov third hypothesis universal constant $\mu$ in (5) for high Re values should be 0.4-0.5.  For higher order spectra $E_n$, with integral $u^n$, Kolmogorov's theory predicts $E_n \sim (\ )^{2n/3} k^{-(2n+3)/3}$, compared to $E_n \sim (u_o\ )^{n/2} k^{-(2n+2)/2}$ using Kraichnan's variables. Neither theory agrees for n>1 with the experimental result of Van Atta and Wyngaard (1975) showing that $E_n \sim k^{-5/3}$ for all n in the range 1-9 at large k.  The reason appears to be that adiabatic, or sweeping, interactions between the large scales of turbulence have time to affect the small scales of higher order spectra, as they do for the dissipation spectrum, but not for the kinetic energy spectrum with n=1.  This approach has been applied by Nelkin and Tabor (1990), and apparently supplies (A. Bershadskii, personal communication) a justification for the Van Atta and Wyngaard (1975) assumption that the measured forms of higher order spectra represent passive scalar mixing, depending on the dissipation rates $\ _{2n,2n+1}$ of $u^{2n}$ and $u^{2n+1}$ just as the spectrum $E_T \sim \ ^{-1/3} k^{-5/3}$ of temperature fluctuations depends on the dissipation rate $\ $ of the temperature variance.

## 3.  Oceanic turbulence

Most of the ocean is so strongly stratified that turbulence is inhibited.  Turbulence forms very rarely, and after it forms it is rapidly converted to buoyancy dominated or Coriolis force dominated classes of motion that are not turbulent by our definition.  Energy and momentum is continually exchanged back and forth between turbulence events and a wide variety of different classes of motion such as internal waves, currents, and "two dimensional turbulence" motions that are neither two dimensional nor turbulence.

The physical situation is similar to that in astrophysics, where many important phenomena involve "plasma turbulence" oscillations and drift currents that are dominated by electromagnetic forces.  Kadomtsev (1965) makes this distinction clear in the Introduction to his classic book on plasma turbulence, where he defines plasma turbulence as "the motion of a plasma in which a large number of collective degrees of freedom are excited.  Thus, when applying the term 'turbulence' to a plasma, it is used in a broader sense than in conventional hydrodynamics".

Unfortunately, it has become conventional in oceanography to treat as turbulence virtually any random motion or scalar fluctuation.  More than fifty years ago Sverdrup et al. (1942, p90) declared that, "In nature, laminar flow is rarely or never encountered, but, instead, turbulent flow, or turbulence, prevails."  Internal wave and Coriolis wave and eddy motions may be quite important to the diffusion and dissipation of the fluid properties in the ocean, and may have other properties in common with turbulence such as random behavior and enhanced diffusion of scalar properties, but they are quite different classes of fluid motions, governed by different equations and exhibiting different behaviors, and should not be called turbulence without qualification. Just as with weak interactions in plasma turbulence, these wave motions require long periods of time to interact compared to their oscillation periods, in contrast to fluid mechanical turbulence where such periods are identical. Internal waves, modons, heatons, vortical modes, and fossil turbulence are the fool's gold of ocean microstructure. All that wiggles is not turbulence. Treating remnants of previous active turbulence as representative samples of the turbulence process leads to the dark mixing paradox, Gibson (1991b), where dark mixing (a concept due to Tom Dillon) is unobserved mixing in the ocean that must exist somewhere to explain the fact that the ocean is mixed, just as dark matter is unobserved matter in galaxies and the universe that must exist to explain why galaxies and the universe are gravitationally bound.

During the last twenty years, numerous ocean microstructure measurements have revealed that only a small volume fraction, about 5%, of most ocean layers are occupied by patches of temperature and salinity microstructure. Such microstructure is the signature of turbulence and turbulent mixing. The microstructure appears in patches and clusters of patches, often associated with frontal structures. The fronts are boundaries of eddy-like horizontal features related to turbulence through hysteresis-like fossil turbulence phenomena, Gibson (1980, 1982, 1986, 1991abc). Rare turbulence events appear at small scales and cascade to large as described in Section 2. Vertical eddy motions are converted to non-turbulent "wave" motions by the damping action of buoyancy forces. The damping takes place when the inertial-vortex forces of turbulence, proportional to $V^2(L)L^2$, are balanced by buoyancy forces $N^2L^4$. This occurs at the Ozmidov length scale $L_R \approx (\varepsilon/N^3)^{1/2}$ found by setting these forces equal, where $V(L) \approx (\varepsilon L)^{1/3}$ for turbulence by Kolmogorov's second hypothesis. Powerful "fossils" may trigger a cascade of other waves and further turbulence events that extract energy from shear flows in a much larger volume of fluid that occupied by the initial event. The stratification, or Väisälä, frequency is defined as $N \equiv (g \partial \rho \partial z/\rho)^{1/2}$, where g is gravity, $\rho$ is density, $\partial \rho \partial z$ is the ambient vertical density gradient, and z is down. The maximum overturn scale of turbulence $L_T$ in a stably stratified fluid is set by $L_R$. For a given ambient N, a turbulence patch can grow until $L_T$ is about $0.6L_R$ where damping begins and vertical growth ceases. Horizontal turbulence growth can continue, however, if horizontal velocity differences exist to supply the energy, up to a maximum at the Hopfinger scale $L_H \approx (\varepsilon/f^3)^{1/2}$ that emerges by setting inertial-vortex forces equal to Coriolis forces, where f is the vertical Coriolis parameter $2\Omega \sin(\theta)$, $\Omega$ is the angular speed of the earth, and $\theta$ is the latitude. When the turbulence is driven by a buoyancy flux q rather than shear, the horizontal scale of turbulence is the Fernando scale $L_F \approx (q/f^3)^{1/2}$. Assuming these driving forces are independent of latitude, one would expect the range of turbulence length scales, and the intermittency factors of turbulence dissipation from equation (5), to decrease from a maximum at the equator to a minimum near the poles, and this appears to be consistent with observations, Baker and Gibson (1987), where maximum values of $\sigma^2_{\ln \varepsilon_r}$ of 7 were observed near the equator and smaller values of 3-5 at higher latitudes.

## 3.1. Hydrodynamic phase diagrams

An important tool for classifying oceanic microstructure patches according to hydrodynamic state is the hydrodynamic phase diagram, as shown in Figure 3. Three rates of viscous dissipation are needed for a microstructure patch to determine its hydrodynamic state with such a diagram: $\varepsilon$, $\varepsilon_F$, and $\varepsilon_o$; the actual, completely fossilized and the beginning fossilization values, respectively. The actual $\varepsilon$ value in a microstructure patch can be measured directly or estimated from a fit to universal spectral forms, or both. The dissipation at complete fossilization is

$$\varepsilon_F = 30 \, \nu N^2 \qquad (7)$$

according to a derivation in Gibson (1980), where $\nu$ is the kinematic viscosity and N = $[(g/\rho)\ \partial\rho / \partial z]^{1/2}$ is the local stratification frequency averaged over a vertical distance larger than the patch. By definition, active turbulence exists in the microstructure patch if $\varepsilon$ is greater than $\varepsilon_F$. Numerous comparisons to laboratory and field data have confirmed the validity of the universal constant given in (7), as summarized in Gibson (1991d). A recent demonstration of the fossilization process is in the stratified wake of flow past a cylinder in a water channel, by Xu et al. (1995).

The dissipation rate at fossilization $\varepsilon_o$ can be estimated from the Cox number $C_o$ at the beginning of fossilization according to the Gibson (1980) fossil turbulence model

$$\varepsilon_o = 13DC_oN^2 \qquad (8)$$

where the Cox number C is defined as the mean squared gradient of a scalar quantity mixed by turbulence such as temperature T, divided by the squared mean gradient

$$C = \overline{(\nabla T')^2} / (\overline{\nabla T})^2 \qquad (9)$$

and averaging is over the patch. The Cox number C is a good estimator of the lower bound of the value at beginning of fossilization $C_o$ because mixing in a fossilized turbulence patch continues at high levels for a long period after most of the turbulence has been damped. Another estimator for $\varepsilon_o$ is the vertical extent of an isolated microstructure patch, since this is $1.5L_{R_o}$ according to laboratory measurements of a stratified jet by Gibson (1991d, fig. 9, p12,562). Thus

$$\varepsilon_o = 0.44 \, L_P{}^2N^3 \qquad (10)$$

can be estimated from the vertical extent of a patch $L_P$ and the ambient N.  Lower bounds for $L_P$ and  $_o$  can be conveniently estimated from the maximum Thorpe displacement scale $L_{Tmax}$ determined by reordering the vertical profile of temperature to be monotonic.

Figure 3 is a plot of ( / $_o$)$^{1/3}$ versus  / $_F$, and can be interpreted as a plot of the normalized Froude number versus normalized Reynolds number for microstructure patches.  Froude number Fr is usually defined as the square root of the ratio of inertial to buoyancy forces

$$Fr = U/LN \tag{11}$$

where U is a characteristic velocity on scale L in a stably stratified fluid with frequency N.  The Froude number of the microstructure patch on scale L divided by the Froude number at fossilization follows from Kolmogorov's second similarity hypothesis and (11)

$$Fr/Fr_o = ( / _o)^{1/3} \tag{12}$$

where U for turbulence is proportional to ( L)$^{1/3}$ by hypothesis.

Reynolds number Re is usually defined as the ratio of inertial forces to viscous forces

$$Re = UL/ \tag{13}$$

where   is the kinematic viscosity of the fluid.  In a microstructure patch after fossilization begins, the largest scale turbulence is at the Ozmidov scale $L_R = ( /N^3)^{1/2}$, with Re given by

$$Re = ( L_R)^{1/3}L_R/ = / N^2 \tag{14}$$

so

$$Re/Re_F = /30 N^2 \tag{15}$$

from (14) and (7).  Note that $Re_F = 30$.

As shown in Fig. 3, a turbulence event in a stably stratified fluid body like the ocean begins, as usual, at small scales, at critical Froude and Reynolds numbers near point 1, and evolves into the fully active turbulence regime with larger and larger scales to point 2 where buoyancy forces again limit the largest vertical scales of the turbulence.  After fossilization begins at  point 2, the patch evolves along a line of slope +1/3 to points 3 and 4 as   decreases but  $_o$ and  $_F$ remain nearly constant (the patch height $L_P$ does not decrease as the patch "collapses").  For the 22 m vertical equatorial undercurrent patch reported in the shear layer above the core by Hebert et al. (1992), shown by the circle, this occurs with a Reynolds number ratio of over 10$^6$, and is confirmed by other (more fossilized) patches with even larger overturn scales, up to 30 m, by Wijsekera and Dillon (1991), triangles, and Peters et al. (1994), square.  Such patches are quite rare, but totally dominate the average dissipation rates and vertical diffusivities through the upper several tens of meters in the upper equatorial ocean.  For example, if the average dissipation rate for the strongly stratified upper layers of the equatorial ocean are less than  $_F$, as reported by Peters et al. (1994, 1995), one 30 m patch with $_o = 3 \times 10^6$ $_F$ will more than double the measured average in $10^8$ m of water sampled. However, at 1 m/s sampling speed this would require about 3 years of continuous sampling in the layer to encounter such a patch. The partly fossilized turbulence remnants of such patches prove that they were once there and fully active, and prove that sampling strategies that fail to account for extreme intermittency effects are likely to underestimate turbulence dissipation and diffusion rates by truly enormous values, Gibson (1990).

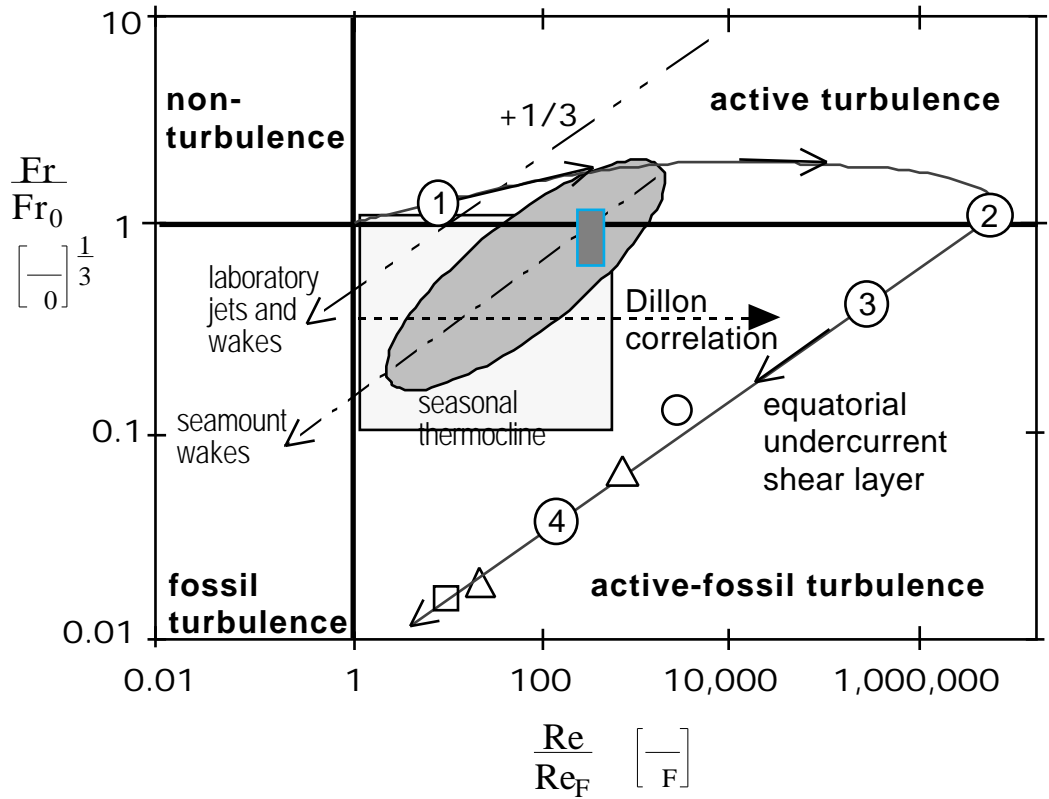

**Figure 3.** Hydrodynamic phase diagram for representative stratified microstructure data sets. Arrows with slope +1/3 show the decay trajectory of a fossilizing microstructure patch. The Dillon correlation $L_R = L_{T rms}$ is shown by the horizontal dashed arrow. Large equatorial patches with density overturns of twenty to thirty meters have been observed by Hebert et al. (1992) ◯, Wijesekera and Dillon (1991) △, and Peters et al. (1994) ☐, indicating $_o$ values in previous actively turbulent states more than a million times $_F$ and thousands to hundreds of thousands times the measured values of $\cdot$ . Such patches in their fully active turbulence regime dominate the space-time average turbulence dissipation rates and vertical fluxes through the layer.

These powerful events are probably caused by persistent shears generated at fronts at the boundaries of horizontal turbulence eddies on hundreds of kilometer scales permitted by the lack of Coriolis forces at the equator.  Strong internal waves termed "fossil turbulence waves", Gibson (1991b), are produced by these turbulence events which irradiate and activate turbulence in all directions in surrounding volumes several orders of magnitude larger than the patch volume, contrary to the common assumption that turbulence in the ocean is always caused by breaking internal waves, Gregg et al. (1993).  Gibson (1996b) suggests that the strong intermittency of internal wave shear reported by Gregg et al. (1993) is evidence that most of the small scale internal wave energy of the ocean may be produced by turbulence, rather than the other way around.

Microstructure data from the seasonal thermocline generally falls in the larger square box shown in Fig. 3, with much smaller values of   and  $_o$.  Microstructure data from laboratory jets and wakes, Gibson (1991d), have the advantage of more precise measurement, but have even smaller values of   and  $_o$.  Evolution of fossilizing turbulence in laboratory wakes and jets, and in the wake of Ampere Seamount (small rectangle), both follow the characteristic decay line with slope +1/3, Gibson et al. (1994).

Turbulence and fossil turbulence in rotating non-stratified, and rotating stratified turbulence may be classified according to hydrodynamic state, as in Fig. 3, by hydrodynamic phase diagrams proposed by Gibson (1991d) using normalized Rossby versus normalized Reynolds numbers, and normalized Rossby versus normalized Burger numbers, respectively.  However, very little relevant data exists for comparison using such diagrams.  Hydrodynamic phase diagrams may also be used to identify fossil turbulence and the degree of fossilization produced by other constraints to turbulence such as electromagnetic forces and self-gravitation forces.

## 4.  Atmospheric turbulence

Near the surface of the earth in daylight hours the atmosphere is maintained in a turbulent state by surface heating up to the "inversion" depth proportional to $L_R$ where buoyancy forces begin to dominate.  Evaporation from the ocean, seas, lakes and land supply moisture of the turbulent weather patterns that distribute water crucially needed for agriculture and many other human uses.  Consequently, every aspect of atmospheric turbulence is subject to intense scrutiny to extract clues to improve weather prediction methods, or simply to improve basic understanding of measurement methods and transport processes.  Approximately 300 papers per year on turbulence in the atmosphere were published during the last five years, mostly dealing with these practical aspects, compared to less than 80 per year on turbulence in the ocean.  As in oceanography, efforts are made to model turbulence by numerical methods, for example Hunt et al. (1991) for complex terrains.  Large, McWilliams and Doney (1994) review models of upper ocean mixing processes, and compare to atmospheric methods.  Mason (1994) reviews the method of turbulence-closure by means of large-eddy simulation. Frenzen and Vogel (1992) review experimental uncertainties in measurements of the turbulent kinetic energy budget for the atmospheric boundary layer.

Basic research on turbulence in the atmosphere is hampered by sensor response limitations on the high speed platforms available above the inversion layer. Consequently, more is known about stratified turbulence in the ocean than in the atmosphere because dropsonds and towed bodies can be deployed from ships. Microconductivity probes, thermistors and shear probes on these oceanic platforms can resolve both viscous and temperature dissipation rates while their atmospheric turbulence counterparts on aircraft generally cannot, as shown by Payne et al. (1994), Fuehrer et al. (1994), Friehe and Khelif (1991). Continuing efforts have focused on calibration of airborne microwave scatterometers for deployment on satellites as surface stress and heat flux sensors versus airborne and surface measurements, Weissman et al. (1994).

## 5. Galactic and extra-galactic turbulence

A flood of new information about the galaxy and universe is emerging from a variety of instruments in space and computers on earth. Color photographs of galaxies in deep space at the beginning of time, galaxies in collision, and stars in formation in gas clouds of our galaxy, are transmitted from the Hubble Space Telescope at high frequency and with stunning clarity. Equally remarkable radio, gamma, ultraviolet, and infrared band telescopic images have been detected from space and the Earth surface, and must be integrated into a new world view. Silk (1994) refers to this as the "golden age of cosmology", and the same might be said of astronomy and astrophysics. What role did turbulence play in the formation of the structures of the universe and galaxy? When did the first turbulence appear? Today we have some hope of answers.

### 5.1 Length scales of self-gravitational condensation

Turbulence is a crucially important phenomenon in the past evolution of the galaxy and its present components, and may have played some part, probably secondary to viscosity, in the evolution of the hot, superviscous plasma universe that existed before the condensation of galaxies or stars. Few comments about turbulence or viscosity are included in recent references on cosmology such as Silk (1994), Padmanabhan (1993), Kolb and Turner (1990) and Peebles (1993). Only Jeans instabilities on acoustic waves in ideal fluids are assumed relevant to the self-gravitational formation of structures in the universe by these authors. A different view is that self-gravitational condensation is limited primarily by viscous and turbulence forces, Gibson (1988, 1996a). We shall question the general relevance of Jeans instability to self-gravitational condensation in the following.

Possibly turbulence is not mentioned in the modern texts because strong primordial turbulence is inconsistent with the extreme isotropy observed in the cosmic background radiation (CBR). The radiation is shown by the 1989 COBE (cosmic background explorer) satellite to have temperature, and therefore velocity, fluctuations of only about $10^{-5}$. Viscosity is not mentioned, possibly because attempts by Misner (1968) and Weinberg (1972) to explain the large entropy of the universe indicated by the CBR data as due to viscous dissipation failed by factors of order $10^7$, Gron (1990), Brevik and Heen (1994). Caderni and Fabbri (1977) conclude the entire microwave background may have originated from extreme viscous dissipation heating during the lepton era ($10^{-5}$ to 24 s with T $10^{12}$ to $10^9$ K). Brevik and Heen (1994) suggest that an impulsive bulk viscosity of $10^{59}$ kg m$^{-1}$ s$^{-1}$ is needed at the end of the inflation era to produce the large observed entropy of the universe, and that the dynamical viscosity µ should be larger than by a factor of at least $10^{16}$. The density at the end of the inflation era, $10^{-33}$ s, was about $10^{51}$ kg m$^{-3}$, giving a kinematic viscosity of $10^{23}$ m$^2$ s$^{-1}$ and a Reynolds number Re$_{max}$ (taking U = c and D = 2 m) of $6 \times 10^{-31}$ or less. These are crude estimates, but they point to a decidedly laminar initial condition for the universe. This laminar condition of uniform rate-of-strain t$^{-1}$ (the Hubble "constant") probably persisted for thousands of years until the universe cooled sufficiently for matter to dominate radiation. The first viscous self-gravitational condensations of the plasma may have produced the first turbulence, but it was weak and rapidly damped as shown by the COBE measurements. The maximum velocity differences on scales of 36° over the full sky are cx$10^{-5}$, or 3000 m s$^{-1}$, where c is the velocity of light. However, scales of 36° correspond to length scales of $5 \times 10^{25}$ m, or $3 \times 10^{22}$ at that time, which are larger than the Hubble scale of causal connection at that time ct = $3 \times 10^{21}$ m by a factor of ten. Therefore the maximum turbulence dissipation rate was only about $300^3/3 \times 10^{20}$ = $10^{-13}$ m$^2$ s$^{-3}$, millions of times less than typical oceanic values or values in the present galaxy, Gibson (1991a).

Chandrasekar (1950, 1951ab), Ozernoy (1972, 1974, 1978), von Weizsäcker (1951), and Silk and Ames (1972), among others, all assume strong turbulence dominated the early universe and that turbulence had a determining influence on structure formation. Gamov (1952) includes extensive discussion of the role of primordial turbulence in structure formation, which, as mentioned, Gamov (1954) suggests as a form of "fossil turbulence". However, beginning in 1965 observations of the extreme uniformity and high intensity of cosmic background radiation, red shifted by a factor of a thousand from the instant of plasma neutralization about 300,000 years after the Big Bang, gradually revealed serious problems with turbulence models. Zel'dovich and Novikov (1983) suggest from these earliest T and u isotropy measurements of the CBR that the Ozernoy whirl theory of turbulent structure formation requires about two orders of magnitude larger levels of turbulence than observed, and therefore cannot be correct.

Silk (1994, p179) assumes that self-gravitational instability and gravitational condensation must occur at scales larger than the Jeans (1902, 1929) length scale L$_J$, first derived by the English astrophysicist Sir James Jeans for an infinite uniform ideal fluid with density and pressure p

$$L_J = V_s/(\rho G)^{1/2}, \tag{16}$$

where $V_s = (p/\rho)^{1/2}$ is the velocity of sound, and G is Newton's constant of gravitation $6.7 \times 10^{-11}$ $m^3$ $kg^{-1}$ $s^{-2}$, and that condensation at smaller scales is impossible. Jeans assumed an infinite, homogeneous, ideal fluid, with small acoustic perturbations of pressure and density. The Jeans analysis is repeated by Zel'dovich and Novikov (1983, pp. 240-244), Peebles (1993, pp. 108-119), Padmanabhan (1993, Ch. 4) with general relativity corrections, Tegmark (1994), Bonazzola et al. (1992), Binney and Tremaine (1987, pp. 287-296) with extensions to non-collisional stellar fluids, and elsewhere, and has been widely accepted as the one and only criterion for self-gravitational condensation. However, such ideal fluid condensation models are highly questionable since actual condensation processes are intrinsically nonlinear and irreversible. Furthermore, density fluctuations can exist without pressure fluctuations. An equivalent (invalid) derivation is to assume that pressure forces $F_P \approx pL^2$ are just balanced by self-gravitational forces $F_G \approx G \rho^2 L^3 L^3/L^2$ at the Jeans length $L_J$ taking $V_s = (p/\rho)^{1/2}$. This is an inappropriate model because the only pressure force available for an ideal fluid starting from rest arises from Bernoulli's equation and has the wrong sign. Such ideal fluid pressures would neither inhibit nor prevent self gravitational collapse. Given any nucleus of density, condensation can take place (given a sticking mechanism like van der Waals forces, or collisions with a nucleus) at scales smaller than $L_J$ if the inertial-vortex forces of turbulence and viscous forces permit, and cannot take place at scales larger than $L_J$ if either the inertial-vortex or viscous forces are larger than self-gravitational forces at such scales.

### 5.1.1 The Jeans (1902, 1929) acoustic self-gravitational length scale

Because Jeans instability is so widely accepted as a description of self-gravitational condensation, it is useful to examine its origins. Jeans (1929, p. 345-355) believed that "spiral nebulae" such as the Andromeda galaxy were composed of gas rather than stars, and were the first step in a top-down condensation sequence from uniform gas to nebulae, stars, planets and moons. His idea of "gravitational instability" was that nearby density perturbations (always accompanied by compensating pressure perturbations) would propagate as sound waves until dissipated by viscosity to heat, but that widely separated density perturbations would grow in kinetic energy "indefinitely" (p345), "the instability entering through displacements in which condensations and rarefactions occur in pairs at sufficiently distant points". He then proceeds with a linear perturbation analysis using Euler's momentum equations including only the pressure and self-gravitational terms, leading to a critical wavelength $\lambda_o$ defined by

$$\lambda_o^2 = (dp/d\rho)(\pi/G\rho) \tag{17}$$

where density fluctuations with wavelengths larger than $_o$ were unstable and would grow indefinitely, and those smaller would not grow. From (16) and (17) we see that $L_J = {}_o/{}^{1/2}$. The predictions of Jeans' theory are illustrated in Figure 4 for three sound waves with wavelengths less than, equal to, and greater than the Jeans critical wavelength.

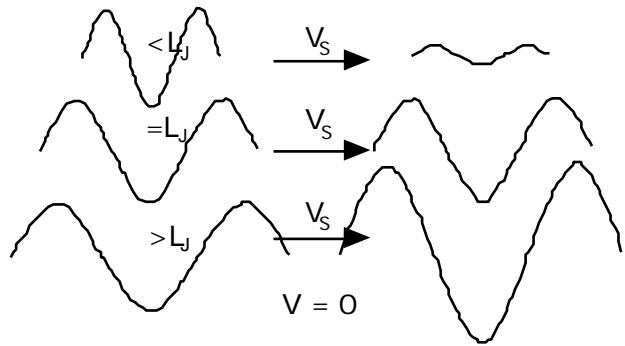

**Figure 4.**  Self-gravitational condensation on density perturbations caused by sound waves of different wavelength and frequency in a stagnant medium, from Jeans (1902, 1929).  The speed of sound is $V_S$, and the critical wavelength for growth is $L_J$ from (16).

The problem with Jeans' conclusion is that he does not consider the possibility of a condensation nucleus; that is, a positive density perturbation in isolation, with no compensating pressure perturbation to cause it to disperse.  Jeans assumed pressure was a function of density only (p346).  Any nucleus placed in a large body of stationary gas will exert an uncompensated gravitational force on its surroundings, which will then begin to collapse on the nucleus until a resistance is encountered, as shown in Figure 5.

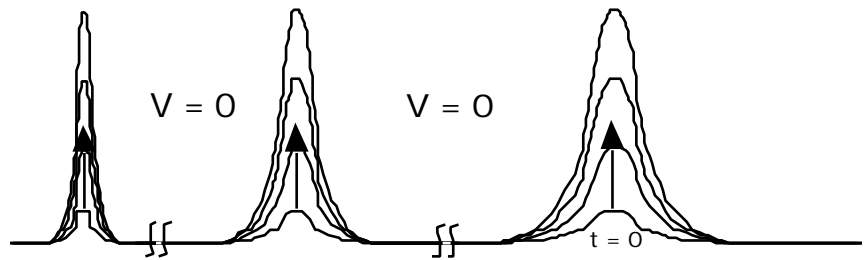

**Figure 5.**  Density perturbations in a stagnant fluid with density    will grow by self-gravity at all scales (assuming the condensing particles can "stick").

No resistance to collapse is available from Euler's equations if the initial pressure is constant.  Viscous, turbulence, or other forces may arise, but are not described by (17).   Jeans' analysis shows that the time required for sound to propagate out of a region of size $_o$ is less than the time of gravitational collapse $(\rho G)^{-1/2}$ so such sound waves will not grow, but that sound waves with wave lengths larger than $\lambda_o$ cannot propagate away fast enough, so their density perturbations will grow indefinitely. This does not prove that gas clouds smaller than $L_J$ will not condense: they will, especially if they contain nuclei and lack turbulence.

### 5.1.2  Viscous and turbulence self-gravitational length scales

Gibson (1988) considers condensation on a nucleus M from a laminar fluid of viscosity $\nu$ with rate-of-strain $\gamma$, and derives a minimum length scale of condensation $L_g$,

$$L_g = GM/\nu\gamma \qquad L_K \qquad (18)$$

where $L_K$ is the Kolmogorov scale, viscous forces dominate gravitational forces on scales L smaller than $L_g$ and gravitational forces dominate viscous forces on larger scales. For a turbulent fluid, the minimum scale of condensation is $L_{g'}$,

$$L_{g'} = (GM)^{3/5}/(\varepsilon)^{2/5} \qquad L_K \qquad (19)$$

where the viscous dissipation rate of the turbulence is $\varepsilon$. The speed of sound and the Jeans length scale $L_J$ are irrelevant to this condensation process.

Starting from a gas cloud of density $\rho$ with density fluctuations $\rho'$, the turbulent Schwarz radius, Gibson (1996a)

$$L_{ST} = \varepsilon^{1/2}/(\rho G)^{3/4} \; ; L_{ST} > L_K \qquad (20)$$

is derived by equating the inertial vortex forces $F_I \approx \rho (\varepsilon L)^{2/3} L^2$ that would disrupt gravitational condensation of a turbulent fluid to the self gravitational forces $F_G \approx \rho^2 G L^4$ that drive the condensation, where $\varepsilon$ is the dissipation rate and $L_K$ is the Kolmogorov scale. $L_{ST}$ is closely analogous to the Ozmidov scale $L_R = (\varepsilon/N^3)^{1/2}$ for stably stratified turbulence. The quantity $(\rho G)^{-1/2}$ represents the characteristic time of self-gravitational condensation, just as $N^{-1}$ represents the characteristic period of oscillation in a stably stratified fluid. Thus, stars form more rapidly where the fluid density is large, but at very large scales when the fluid is very turbulent, from (20). The gravitational-inertial-viscous scale

$$L_{GIV} = (\nu^2/\rho G)^{1/4} \qquad (21)$$

where all three forces are equal is closely analogous to the buoyancy-inertial-viscous scale $L_{BIV} = (\nu/N)^{1/2}$ of complete fossilization in a stably stratified fluid, Gibson (1991a).

For laminar flows, viscous forces will disrupt the condensation on length scales smaller than the viscous Schwarz radius

$$L_{SV} = (\nu\gamma/\rho G)^{1/2} \; ; L_{ST} < L_K, \qquad (22)$$

where   is the kinematic viscosity of the fluid and   is the rate-of-strain, Gibson (1996a).  $L_{SV}$ is derived by equating viscous forces $F_V$    μ $L^2$ to $F_G$      $^2GL^4$.  Once condensation begins, stable stratification can accelerate the process by damping turbulence in the condensing fluid in the absence of embedded angular momentum.  The viscous and turbulent Schwarz radii are equivalent to $L_g$ and $L_{g'}$ of (18) and (19), respectively, if M =   $L^3$ [the terminology "Schwarz radii" is proposed in Gibson (1996a) in honor of the late William H. Schwarz].  Condensation will occur in clouds of scale L as long as the convective velocities permit, as shown in Figure 6.

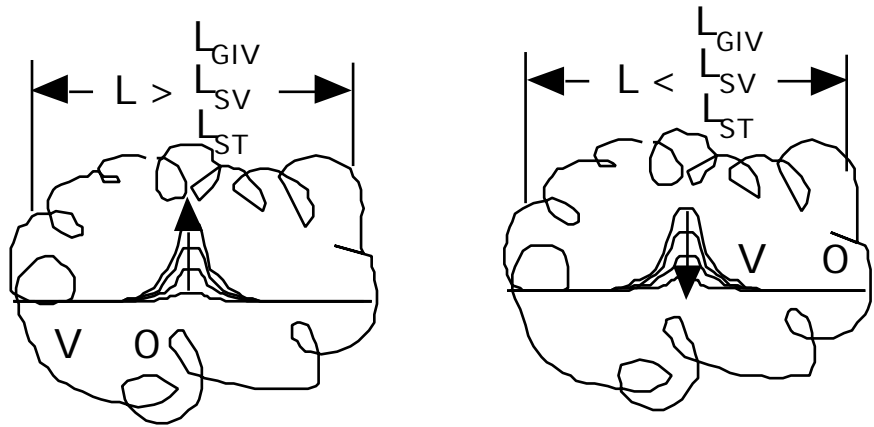

**Figure   6.**  Self-gravitational condensation in gas clouds of scale L is possible (left side) or impossible (right) depending on a comparison of L with the Schwarz gravitational scales.

$L_{ST}$ and $L_{SV}$ represent minimum scales of self-gravitational condensation, so in principle condensation can occur with the same time scale (  G)$^{-1/2}$ for larger scales if the extent of fluid with density   is larger than $L_{ST}$ or $L_{SV}$.  However, we can expect that most of the condensation will initially be at the smallest possible scales $L_{ST}$ and $L_{SV}$ because these can occur rapidly on local density maxima, and will not be disrupted by nuclear reactions and concentrations of vorticity and magnetic fields characteristic of the largest scale astrophysical mass condensations.

Where the fluid is turbulent, $L_{ST}$ is likely to be larger than $L_J$, and where the fluid is laminar $L_{SV}$ is likely to be smaller than $L_J$.  Thus the Jeans criterion is likely to overestimate the rate of star formation in galactic disks, where large young stars are caused by the high Mach number pressure waves emanating from the galactic core that snowplow enough material together to permit condensation.  Scheffler, H., & H. Elsässer (1982) have noted that about 50 times less stars appear per year in the Milky Way galaxy disk than are expected from the Jeans criterion, and suggest the possibility that turbulence may act to inhibit the star formation process.

### 5.1.3  Observations

Figure 7 shows star formation in the pressure wave of the Cartwheel galaxy, caused by a collision between one of the two galaxies on the right with the larger galaxy on the left.

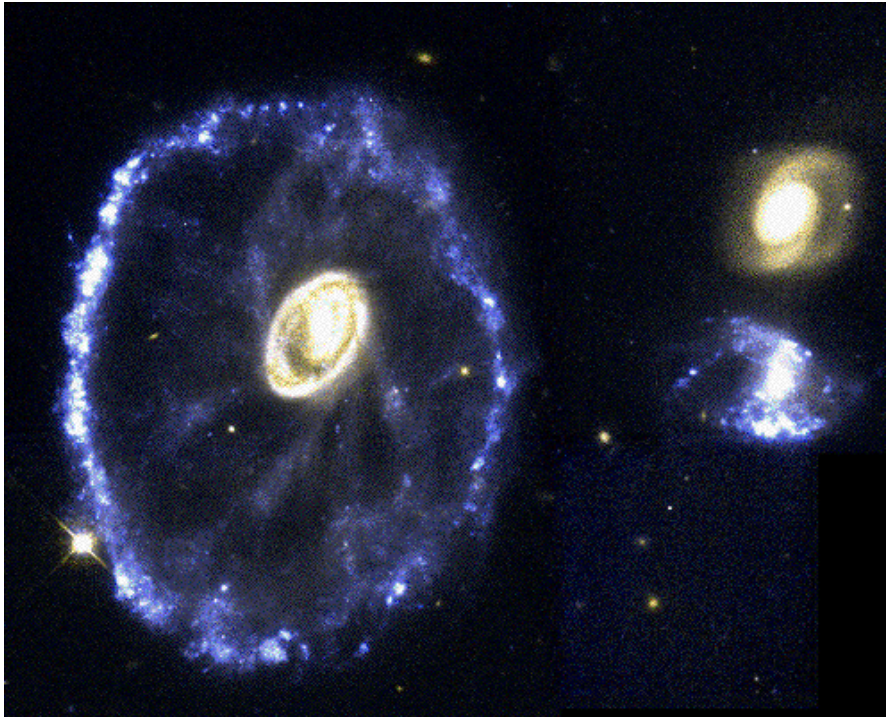

**Figure 7.**  Cartwheel Galaxy.  Several billion huge new stars with thousands of supernovae are produced by the $10^5$ m s$^{-1}$ shock wave from a galactic collision 0.5 billion years previous, that snowplowed material in the galactic disk to the high density and turbulence levels required to such large turbulent Schwarz radii $L_{ST}$ of star condensation.

The Cartwheel Galaxy is at a distance of $5 \times 10^{24}$ m in the direction of constellation Sculptor.  The spectacular cartwheel form is $1.4 \times 10^{21}$ m in diameter, larger than the Milky Way. Such new $10^{32}$ kg blue supergiant stars formed in the "wheel" by the tsunami-like blast wave (velocity    $10^5$ m s$^{-1}$, Mach numbers of order $10^3$) exist for only a few million years before exploding in supernovae that precipitate subsequent star formation as star bursts.  Young blue star births are similarly triggered by pressure waves in the disks of spiral galaxies to form the visible arms, such as the "spokes" of the Cartwheel that are beginning to re-emerge.  Smaller, older, yellow stars are seen in the galaxy core.  The photograph was take by the Hubble Space Telescope, Kirk Borne of ST ScI and NASA, October 16, 1994.  Bright knots on the "cartwheel" are revealed by the HST as millions of stars forming in bursts, not individual star formation as assumed by early astronomers that viewed spiral galaxies as planetary nebulae with stars only in the spiral arms; for example, Jeans (1929).

## 5.2  Structure formation in cosmology

Self-gravitational condensation based on the Jeans instability criterion gives a very different sequence of structure formation in the universe than condensation limited by viscous and turbulent forces.   The speed of sound for ordinary baryonic matter in the hot Big Bang fluid was proportional or equal to the speed of light, so the condensation scale $L_J$ from (16) was too large to permit any condensations smaller than the Hubble scale ct until after plasma neutralization at t about 300,000 years, Kolb and Turner (1994, p. 363).  The Jeans scale after neutralization was still too large for stars to form, so only small galaxy masses and larger could begin to condense, Padmanabhan (1993, p. 135).  The mysterious WIMP material was so weakly interacting that it became electrically neutral much earlier, according to Padmanabhan (1993), and therefore became the first matter to condense.  The primordial WIMP condensation scales were small because the density was large, so Padmanabhan (1993, p. 131-135) suggests these primordial wimp fog (PWF) particles may have become nuclei for the subsequent condensation of stars.  Table 1, from Peebles (1994, p. 611), summarizes the sequence of structure formation expected based on the Jeans instability criterion alone.

| Event | Redshift z | Time, years |
|-------|------------|-------------|
| Gravitational potential fluctuations | 1000 | 500,000 |
| Spheroids of galaxies | 20 | $1.6 \times 10^8$ |
| The first engines for active galactic nuclei | 10 | $4 \times 10^8$ |
| The intergalactic medium | 10 | $4 \times 10^8$ |
| Dark matter | 5 | $10^9$ |
| Dark halos of galaxies | 5 | $10^9$ |
| Angular momentum of rotation of galaxies | 5 | $10^9$ |
| The first 10% of the heavy elements | 3 | $2 \times 10^9$ |
| Cosmic magnetic fields | 3 | $2 \times 10^9$ |
| Rich clusters of galaxies | 2 | $3 \times 10^9$ |
| Thin disks of spiral galaxies | 1 | $5 \times 10^9$ |
| Superclusters, walls and voids | 1 | $5 \times 10^9$ |

**Table 1.  Timetable for Structure Formation by Jeans' Criterion**

The timetable for structure formation based on the viscous and turbulent criteria discussed previously is quite different than that in Table 1.  Superclusters are the first rather than last structures to form, and the WIMP material condenses last, as superhalos of the superclusters, rather than first as PWF particles.  Condensation begins when the viscous Schwarz scale $L_{SV}$ decreases to equal the increasing Hubble scale $L_H = ct$, and this probably took place at a time t comparable to the mass-energy transition at about 10,000 years after the big bang, assuming a flat universe dominated by weakly interactive massive particle, or WIMP, cold dark matter, Habbib and LaFlamme (1994).  Time and space are greatly distorted at such early times, so that it is impossible to ignore effects of general relativity.  Such effects are described by the standard cosmological model of Weinberg (1972).

The standard model was constructed by inserting the energy-momentum tensor of ideal, non-viscous fluids into the Einstein field equations assuming the homogeneous, isotropic metric space of Friedmann-Robertson-Walker (FRW), and various equations of state for ideal cosmic fluids relating the pressure to the density.  Predictions include the scale factor $a(t)$, density $\rho(t)$ and temperature $T(t)$ as functions of time. The Friedmann-Robertson-Walker metric $ds^2 = g_{ij}dx_i dx_j$ is

$$ds^2 = -dt^2 + a(t)^2[dx_i^2] \qquad (23)$$

where $dx_i$ for $i = 1,2,3$ are the three space components, the universe is assumed flat (neither closed nor open), and space-time coordinates are normalized so the velocity of light, Boltzmann, and Planck constants are equal to 1.  The Einstein field equations are

$$R_{ij} - g_{ij}R = -8\pi G T_{ij} \qquad (24)$$

where $R_{ij}$ is the Ricci tensor, R is its trace, $g_{ij}$ is the metric tensor, G is Newton's gravitational constant, $T_{ij}$ is the energy-momentum tensor, and indices i and j are 0, 1, 2, and 3.  The Ricci tensor is composed of first and second derivatives of $g_{ij}$ in a form developed for non-Euclidean geometry by Riemann and Christoffel and adapted by Einstein to preserve Lorentz invariance and the equivalence of inertia and gravitation in mechanics and electromechanics, Weinberg (1972). Curvature effects must be retained in (23) under conditions of extreme gravitation; for example, those associated with black hole dynamics or in the initial stages of the big bang, Thorne (1994). $R_{ij}$ is subject to Bianchi identities relating derivatives of the metric.  These are useful in constructing anisotropic viscous cosmologies.  Solving (24) with different equations of state and $T_{ij}$ for ideal fluids gives

$$a(t) \sim t^{1/2} \ ; \ \rho(t) \sim a(t)^{-4} \sim t^{-2} \ ; T \sim a(t)^{-1} \sim t^{-1/2}; \text{radiation dominated plasma} \qquad (25)$$

$$a(t) \sim t^{2/3} \ ; \ \rho(t) \sim a(t)^{-3} \sim t^{-2} \ ; T \sim a(t)^{-1} \sim t^{-2/3}; \text{ matter dominated plasma} \qquad (26)$$

$$a(t) \sim t \ ; \ \rho(t) \sim a(t)^{-3} \sim t^{-3} \ ; T \sim a(t)^{-1} \sim t^{-1}; \text{ matter dominated gas} \qquad (27)$$

for the radiation and mass dominated plasma epochs, and the present neutral gas epoch. Full discussion of these classical solutions of the Einstein field equations (24) are given by standard cosmology texts such as Weinberg (1972), Peebles (1993), and Padmanabhan (1993). Viscous cosmologies add additional terms to $T_{ij}$ that may cause the time evolution of the scale factor $a(t)$, the density (t), and the temperature $T(t)$ to be somewhat different from (25)-(27). Weinberg (1972, p. 540) provides a table of $a(t)$, (t), and $T(t)$ based on the Einstein equation (24) for an ideal fluid during the first $10^{13}$ s of the universe up to the time of decoupling. This was used by Gibson (1996a) to construct the model for structure formation shown in Figure 8.

As shown in Fig. 8, the evolution of structure in the universe governed by viscous and turbulent forces is quite different from that governed by the Jeans acoustical instability shown in Table 1. The age of the universe is divided into a series of five epochs I-V. The first epoch I. is viscous dominated and has no structure because the causality scale $L_H$ is smaller than all the gravitational condensation scales $L_{GIV}$, $L_{SV}$, and $L_{ST}$. Structure formation begins at a time about 10,000 years when $L_{SV}$ decreases to $L_H$, at temperatures $10^6$ K and scale factor $a(t)$ was about $10^{-6}$ and extends throughout epoch II until the photon decoupling event at 300,000 years. Nested clouds of plasma with a foam-like topology should form during this epoch, reflecting the Zel'dovich "pancake" phenomenon of a preferred axis for gravitational collapse.

Thus, galaxy supercluster and galaxy cluster geometry and masses were fixed at this earliest stage of the universe structure formation, rather than the latest as in Table 1. From (26) and expressions for plasma viscosity in Weinberg (1972) it is possible to extrapolate the condensation size of the largest superclusters of $10^{47}$ kg, Kolb and Turner (1993), to the time of photon decoupling at 300,000 years, giving the final condensation size of about $10^{42}$ kg, Gibson (1996a), the size of galaxies.

A period exists at the end of the plasma era, when all matter becomes neutral gas, during which the Jeans instability may be relevant to self-gravitational condensation. The Jeans mass suddenly drops from $10^{49}$ kg for the plasma (too large for condensation because it exceeds the Hubble scale of causality ct) to $10^{36}$ kg for the hot neutral gas, Weinberg (1977, pp. 173-175), the mass of the globular clusters. Condensations of the gas formed within the $10^{42}$ kg protogalaxy droplets will occur both at the viscous Schwarz mass of $10^{22}$ kg to form primordial fog, but may also occur at the density maxima corresponding to sound waves with wavelengths $L_J$ and longer. The smallest permitted acoustic scale is likely to condense first, and this condensation process at $L_J$ can produce resonant sound waves with this wavelength that can trigger further condensations within the protogalactic gas droplets. Thus some fraction of the protogalaxy droplets will form protoglobular-cluster droplets at $L_J$ during the condensation period ( G)$^{-1/2}$   $4x10^{13}$ s (a million years) following plasma neutralization. With more cooling and turbulence, the Jeans scale $L_J$ decreases as $L_{SV}$ and $L_{ST}$ increase. It becomes irrelevant when $L_J < L_{SV}$, $L_{ST}$.

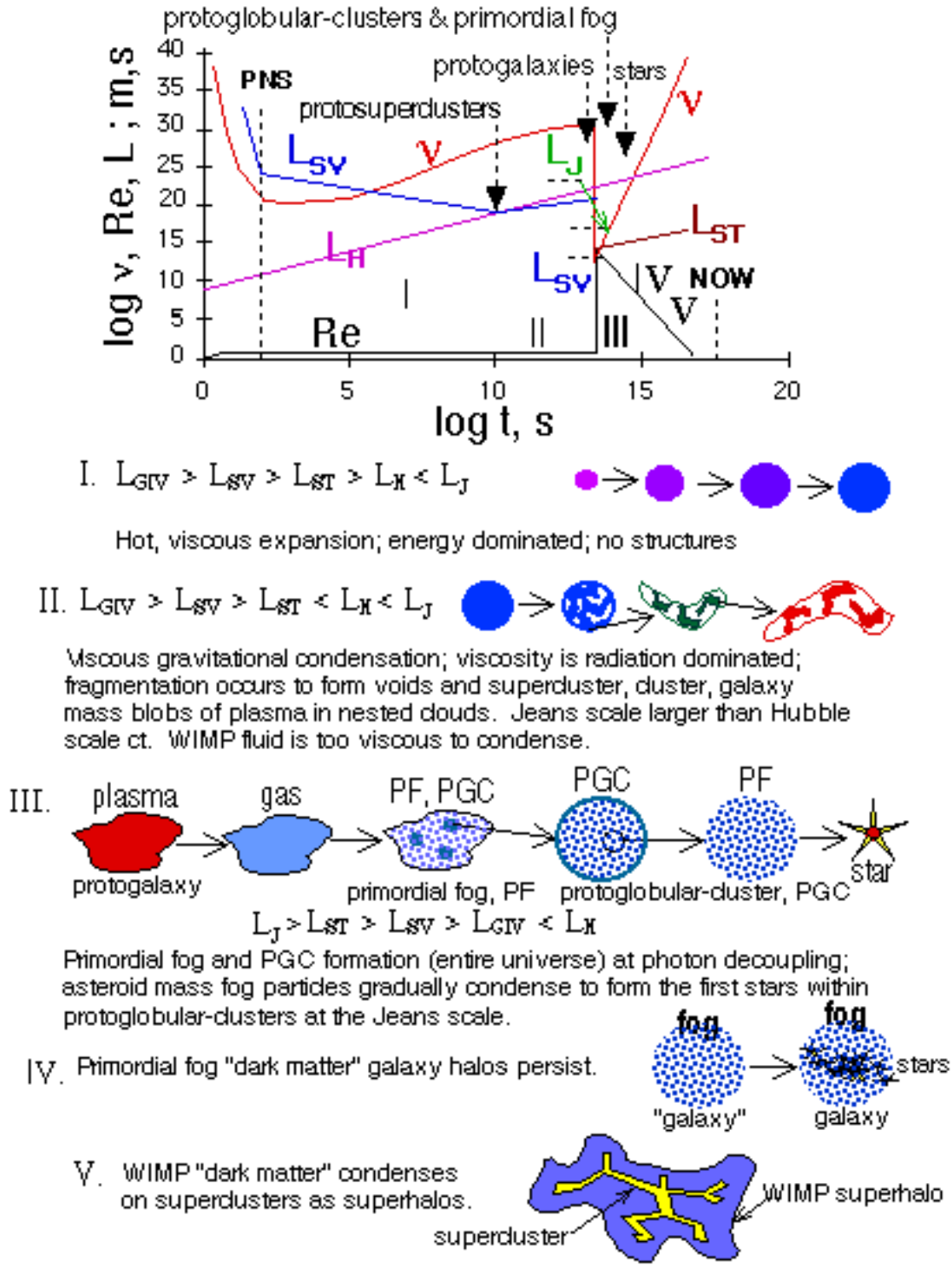

**Figure 8.** Hydrodynamic history of the universe. The top diagram shows the beginning of gravitational condensation when the viscous Schwarz radius $L_{SV}$ decreases to sizes smaller

than the Hubble scale $L_H$.  Sketches below describe time intervals I-V for various epochs of structure evolution discussed in the text.

Plasma neutralization begins epoch III, giving a sudden decrease in viscosity   and in the viscous gravitational condensation scale $L_{SV}$, and a sudden decrease in the size of gravitational condensation, from galaxy mass $L_{SV}^3$  of $10^{42}$ to primordial fog PF mass $L_{SV}^3$   or $L_{ST}^3$  of $10^{22}$ to $10^{26}$ kg.  All the galaxy mass droplets filled with subsolar PF and protoglobular-cluster mass droplets in about $10^6$ y.  These then cooled and compacted during a long period, perhaps $10^8$ y, before the formation of the first star by collisions and accumulation of the primordial fog particles.  Epoch IV is a "bottom-up" period of gradual structure formation within protogalaxies etc. extending to the present, from PF particles to stars to globular clusters of stars, etc.  Epoch V is the gradual condensation of WIMP material on the largest structures to form superhalos on superclusters through the rare collisions necessary for "sticking".  Thus, the model provides two types of dark matter, the baryonic PF particles in the halos of galaxies and the non-baryonic WIMP material in the halos of superclusters and possibly clusters of galaxies.

### 5.3  Observations

Figure 9 shows a globular cluster of small, ancient, stars, whose symmetry reflects the lack of significant turbulence at formation and the extreme homogeneity of their material of construction, consisting of the original uniform size, equally spaced, primordial fog particles.

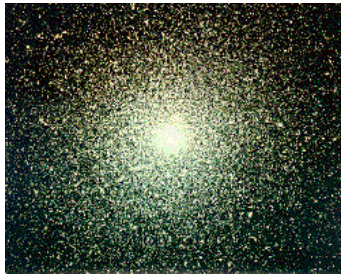

**Figure 9.**  Globular cluster of ancient stars, 47 Tucanae in the Milky Way halo at a distance of $10^{20}$ m: a "fossil non-turbulence" remnant of the gentle early universe.

The density of matter in these most ancient star clusters, termed "globular clusters" because of their spherical symmetry, is about $10^{-17}$ kg m$^{-3}$, close to that at the time of primordial fog formation.  Star ages are estimated to be $1.4 \times 10^{10}$ y by radioactive and chemical methods.  Such globular clusters are concentrated in the cores of most galaxies, including ours, and move in random orbits in their halos.  They contain millions of population II stars consisting only of the primordial mixture of three parts hydrogen and one part helium-4 that was produced at the first nucleosynthesis at 100 s.  Their completely regular shape suggests the epoch of extremely weak turbulence in the universe extended for many millions of years. The existence of huge, spherically symmetric elliptical galaxies such as M87, with mass $10^{43}$ kg, consisting of millions of such ancient globular clusters of population II stars, implies the weak turbulence epoch lasted even longer, perhaps a few billions of years. Dense galaxy clusters have been observed, Dressler (1995, p. 69), to consist mostly of symmetric elliptical galaxies of small ancient stars, while low density galaxy clusters tend to contain only lower density spiral galaxies, rich in metals and larger, bluer, stars like our own turbulent Milky Way. This "Dressler effect" supports the existence of an extremely weak turbulence epoch in the earliest stages of the universe. Galaxies forming after this epoch, more slowly from material in an older universe with smaller density, were subject to disruption by rotational forces of the embedded angular momentum generated by the turbulence of earlier star and galaxy formations, producing galactic disks with spiral arms in widely separated spiral galaxies. By two or three billion years, huge black holes in the centers of the elliptical galaxies could form the engines of the most powerful turbulence in history, the active galactic nuclei. Their light as quasars and their high energy protons at the upper wing of the cosmic ray spectrum can reach us today, many billions of years and light years away.

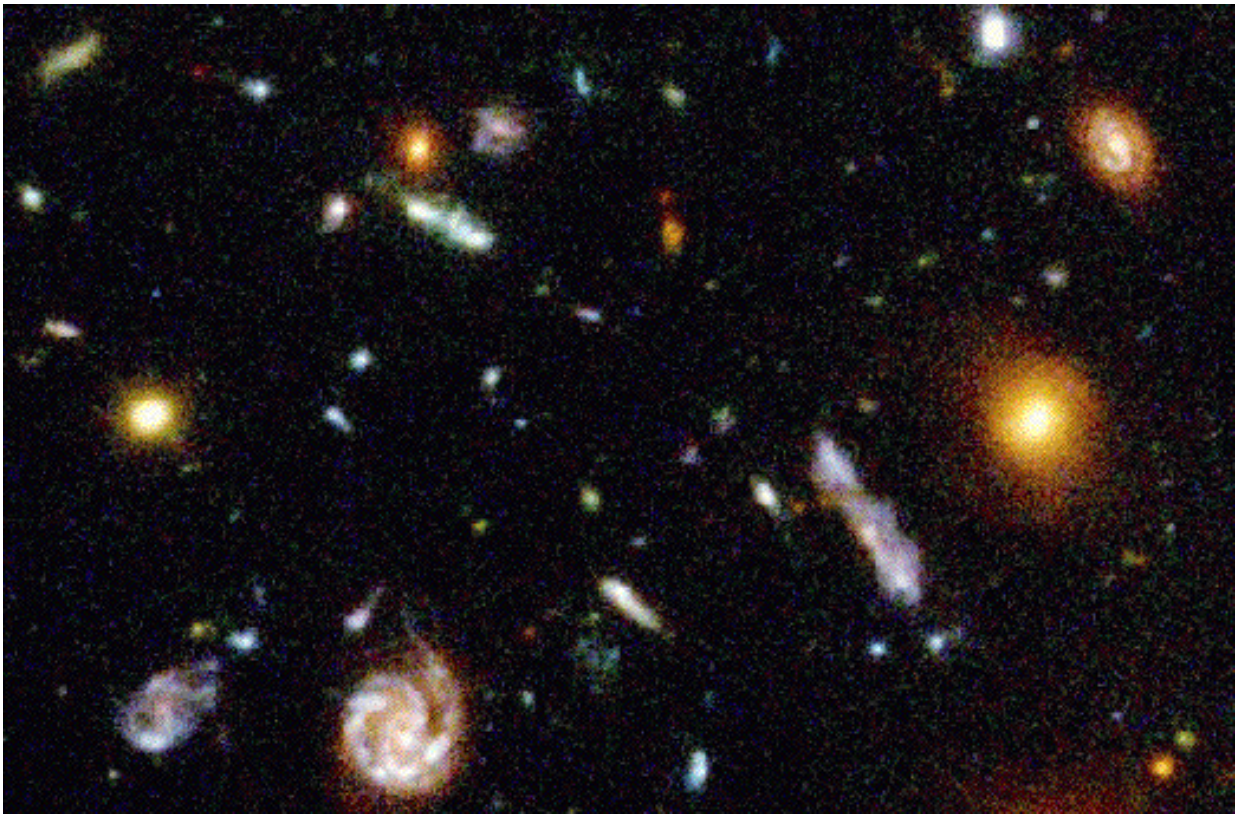

**Figure 10.** Hubble Deep Space galaxies, the most distant ever seen. January 1996 release, Robert Williams, Space Telescope Science Institute, Baltimore, MD.

Figure 10 shows a detail from the recently released Hubble Deep Space photographs (R. Williams, STSci webpage), taken continuously for ten days using four color filters from infrared to ultraviolet for 150 successive orbits of the satellite Dec. 18-28, 1995. This is the deepest and oldest view of ancient galaxies ever taken, and shows an incredible variety of forms and colors. Spherically symmetric (elliptical) galaxies are generally yellow or red indicating their smaller, older stars, formed early in the universe by gradually collecting PF particles before strong turbulence developed to increase $L_{ST}$. Spiral galaxies are white or blue, suggesting turbulence formed by blast waves is forcing larger, high temperature, short lived stars to be formed.

The photograph in Fig. 10 shows galaxies to apparent magnitudes of 30, much fainter than can be seen from surface telescopes, and shows the reason the night sky is dark: there are large gaps between galaxies. Proper statistical analysis of distances, compositions, and ages of galaxies from these photographs should provide valuable comparisons to structure formation models such as presented above. A new infrared camera will be added to the HST in 1997 that will improve the resolution of images of the most ancient, most redshifted, galaxies.

## 5.3 Turbulence in galaxies and intergalactic space

Despite the paucity of matter between the galaxies, some evidence exists of intergalactic turbulence within galactic clusters. Lowenstein and Fabian (1990) infer velocity fluctuations of $10^6$ m s$^{-1}$ from the optical line width of hot gas patches on scales of order $10^{20}$ m, and attribute this to hydrodynamic turbulence. Other authors mention turbulent wakes of galaxies, generation of magnetic fields by turbulence, and the cooling effects of the intercluster gas flow.

The most powerful manifestations of turbulence in the universe are produced by active galactic nuclei and their associated cosmic jets. The nearest is Centaurus A, Burns and Price (1984), which is only $1.5 \times 10^{22}$ m away and covers some 20° of sky. A more distant well studied example is the elliptic galaxy NGC 6251 at a distance of $3 \times 10^{24}$ m, Blandford et al. (1984), with a one-sided straight jet $4 \times 10^{21}$ m long. These are the most luminous objects in the universe, up to $10^6$ times brighter than the average galaxy. The most distant have red shifts of more than 4, velocities 0.9c, and light $4 \times 10^{17}$ s old, or 90% of the estimated universe age. Most of the electromagnetic emission of such quasars comes from their bright central cores. The jets are up to $10^{22}$ m or more long and perfectly straight, or blown back into spiraling radio trails by intergalactic winds that may be over $10^6$ m s$^{-1}$ if the active galaxy is in a dense cluster. Shock waves of the supersonic turbulence in the lobes snowplow the gas to high densities that cause condensation of many huge, bright young stars with masses up to $10^{33}$ kg that condense and explode in times of a few million years or less.

The sources of the jets are central black holes, Thorne (1994). These generate powerful magnetic fields along the black hole axis of rotation, and confined plasma jets. Matter, angular momentum, and magnetic fields are pulled into the black holes by the gravity of some $10^{39}$ kg of mass within: billions of stars out of the trillions existing in the galaxies. The most powerful quasars are formed when "cannibal galaxies" consume "missionary galaxies" through tidal action and accretion by the central nucleus. Turbulence in the accretion disk permits diffusion of part of the angular momentum outward and ohmic dissipation of part of the magnetic fields which would otherwise resist the gravitational collapse and slow the accretion rate. Magnetic forces collimate the jets to narrow angles of less than 3 degrees for the first $10^{20}$ m and suppress hydrodynamic turbulence, although plasma turbulence waves play a role in entraining external gas and reactivating the relativistic electrons, Burns and Price (1984). Figure 11 shows a radio telescope image of an elliptical galaxy with two jets emanating from a source in the active galactic nucleus of an elliptical galaxy.

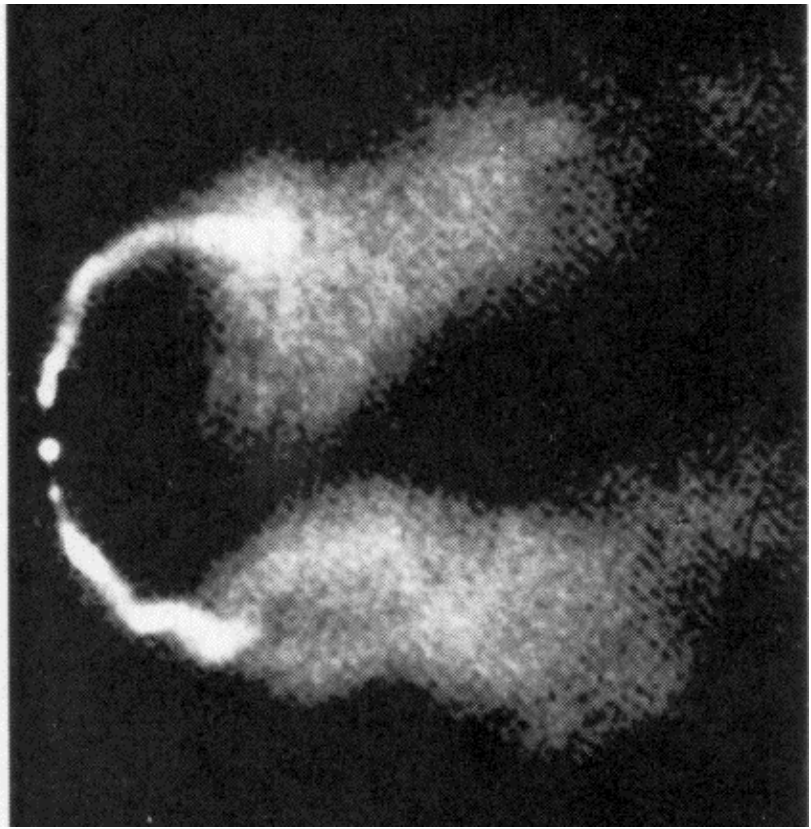

**Figure 11.** "Wingtip vortices" from an active galactic nucleus (galaxy NGC 1265) in the rich Perseus cluster. A spinning central black hole creates plasma beams and turbulent jets $10^{22}$ m long that are blown to the right by intercluster gas flow at $10^7$ m s$^{-1}$. The "smoke" is radio wave emissions from giant stars formed in the turbulent blast waves at huge $L_{ST}$ scales.

Accretion disks form whenever matter condenses under the pull of gravity if the average angular momentum is not zero. Because the matter is fluid, "viscosity" parameters are needed in accretion models, although little is known of the actual fluid dynamic processes. Turbulence slows gravitational condensation in the universe by spreading angular momentum and inhibiting condensation at scales smaller than $L_{ST}$, but ultimately makes condensation possible by providing a means of radial momentum diffusion in accretion disks. From the red/blue shifts in Sc class spiral galaxies with well defined arms such as the Milky Way, Rubin (1984) showed that the radial velocity in the disks was nearly constant rather than decreasing as $r^{-1/2}$ according to Kepler's law for planets, with velocity values about $2.5 \times 10^5$ m s$^{-1}$, which convincingly proves that dark matter must exist in galaxy halos.

Little is known of most galactic turbulence properties. Cosmic jets and accretion disks of black holes may represent examples of extreme turbulence, Blandford et al. (1984). Masers sometimes form around new and old stars, and their Doppler frequency shifts give information about turbulence in the solar wind, Elitzur (1995). Biermann and Strom (1993) show that the spectra of cosmic rays can be explained by turbulence formed by galactic supernovas for the low energy range, and by quasar jets at the highest energies (up to 200 joule per proton!).

## 6. Summary

A review of turbulence in natural fluids suggests it is qualitatively identical to turbulence in the laboratory. Departures from universal similarity hypotheses of Kolmogorov are virtually unmeasurable for low order statistical parameters like spectra of velocity and temperature mixed by turbulence, but are substantial for higher order spectra and dissipation rates, where a sweeping interaction model pioneered by Kraichnan gives more precise predictions.

In natural flows, turbulence is subject to self-gravity, buoyancy, Coriolis, and other force constraints rather than being constrained by laboratory channel walls. Major differences result between laboratory and natural flows because the natural forces couple with the turbulence and often rapidly extract most of the turbulent energy as wave or eddy motions that are not turbulence but have some turbulence properties, and may be mistaken for turbulence. Such waves and eddies preserve information about the turbulence that produced them, and are therefore forms of fossil turbulence. Most ocean microstructure is fossil turbulence. Interpretations of ocean microstructure measurements that fail to recognize this fact underestimate the average dissipation rates of kinetic energy and temperature variance, and therefore the vertical transport rates of heat and momentum, leading to the "dark mixing" paradox of oceanography. Dark mixing is unobserved mixing that must exist to explain the fact that bulk ocean properties are well mixed, just as dark matter is unobserved matter that must exist in galaxies to explain velocity profiles which require additional gravitational forces to be stable besides those of the luminous matter.

Atmospheric turbulence is very well studied for practical reasons. Improvements in numerical modeling of weather and climate can have major economic benefits. Basic research in the turbulence of the atmosphere is hampered by the lack of sensors and platforms that resolve the crucially important dissipation scales over wide enough ranges to include measurements of the macroscales. Oceanic turbulence sensors have the spatial resolution and frequency response required to measure the dissipation rates, but are generally not deployed over adequate space distances and time periods needed to span the full range of oceanic turbulence processes, leading to severe undersampling and underestimates of the true turbulence levels.

A wealth of new information is emerging from space telescopes and improved land based telescopes about turbulence in the galaxy and universe. Interpretation of the evidence based on the classical Jeans criterion for condensation of matter on acoustic waves appears to be generally incorrect and misleading. Comparisons of self-gravitational forces with inertial-vortex forces of turbulence and viscous forces of laminar flow gives new criteria for the minimum condensation scales: the turbulent and viscous Schwarz radii $L_{ST}$ and $L_{SV}$. Application of the new criteria to flows of the early universe suggests that the largest observed structures of the present universe, the superclusters and supervoids, are fossils of the first primordial non-turbulence to emerge from the superviscous fireball of the big bang when the sizes and ages of the universe were $10^{-6}$ to $10^{-3}$ times smaller than those at present. Application to the time when the plasma neutralized at about 300,000 years suggests the formation of a primordial fog of subsolar particles from which all other subsequent structures, such as stars and galaxies, ultimately formed. Simultaneous condensation at the Jeans scale on sound waves in the neutral gas within protogalaxy droplets produces protoglobular-cluster droplets. Present globular star clusters are fossils of the non-turbulence and high density of the early universe. Most of the primordial fog may still remain as asteroid to Earth mass particles in galactic halos as the unobserved dark matter in galactic halos that binds galaxies together by gravity. Another class of dark matter is the weakly interacting (WIMP) material needed to close the universe, that gradually condenses to form superhalos and clusterhalos at large $L_{SV}$ scales corresponding to its superviscosity and the Hubble strain-rate $t^{-1}$ of universe expansion.

## Acknowledgments

The author is grateful for several useful conversations with colleagues about this paper, especially Peter Biermann, George Golitsyn, Evgeny Novikov, Iosif Lozovatsky and Joe Fernando.